\pdfoutput=1
\documentclass[aps,prd,10pt,twocolumn,tightenlines,bibnotes,nofootinbib,superscriptaddress,floatfix]{revtex4-1}
\usepackage{amsmath,amsfonts,amssymb,amsthm}
\usepackage{subfigure}
\usepackage{enumerate}
\usepackage{graphicx}
\usepackage{lipsum}
\usepackage{physics}
\usepackage{bm}
\usepackage[usenames,dvipsnames]{xcolor}
\RequirePackage{color}
\RequirePackage[colorlinks=true
,urlcolor=Maroon
,anchorcolor=Maroon
,citecolor=Maroon
,filecolor=Maroon
,linkcolor=blue
,menucolor=Maroon
,linktocpage=true
,pdfproducer=medialab
,pdfa=true
,pdfusetitle
]{hyperref}
\def\half{{{\frac{1}{2}}}}
\begin{document}

\title{Instantaneous Temperatures \`a la Hadamard:\\ Towards a generalized Stefan--Boltzmann law for curved spacetime}

\author{Aditya Dhumuntarao}
\email{dhumu002@umn.edu}
\affiliation{Perimeter Institute for Theoretical Physics, Waterloo, Ontario, N2J 2Y5, Canada}
\affiliation{School of Physics and Astronomy, University of Minnesota, Minneapolis, Minnesota, 55455, USA}

\author{Jos\'e Tom\'as G\'alvez Ghersi}
\email{joseg@sfu.ca}
\affiliation{Perimeter Institute for Theoretical Physics, Waterloo, Ontario, N2J 2Y5, Canada}
\affiliation{Department of Physics, Simon Fraser University, Burnaby, British Columbia, V5A 1S6, Canada}

\author{Niayesh Afshordi}
\email{nafshordi@pitp.ca}
\affiliation{Perimeter Institute for Theoretical Physics, Waterloo, Ontario, N2J 2Y5, Canada}
\affiliation{Department of Physics and Astronomy, University of Waterloo, Waterloo, Ontario, N2L 3G1, Canada}

\date{\today}

\begin{abstract}
	In the celebrated Unruh effect, we learn that a uniformly accelerating detector in a Minkowski vacuum spacetime registers a constant temperature. Building on prior work, we present a technique based on derivative couplings of the two--point Wightman function and the Hadamard renormalization procedure to define an instantaneous temperature for a massive scalar field, non-minimally coupled to gravity. We find the temperature contains local contributions from the acceleration of the detector, the curvature of spacetime, and the renormalized stress--energy tensor of the field. Our result, which can be considered as a generalized Stefan--Boltzmann law for curved spacetimes, agrees with the familiar expressions found in 4D Rindler, thermal Minkowski, and de Sitter.
\end{abstract}
\maketitle
\section{Introduction}
	The semiclassical treatment of quantum field theory in curved spacetime provides critical insights towards a complete understanding of gravity in the ultraviolet limit. In such a framework, one considers spacetime from the classical viewpoint while the accompanying fields are quantized in an analogous way to Minkowskian quantum field theory. As the notion of a particle propagating in curved spacetime is \textit{nebulous}, the stress--energy tensor, $\textsf{T}_{\textsf{ab}}$, is promoted to an operator and, along with a normalized quantum state $\ket{\Phi}$, the expectation value $\expval{\textsf{T}_{\textsf{ab}}}{\Phi}$ is accepted as the natural coupling in the semiclassical Einstein field equations
	\begin{equation}\label{eqn1:QEFE}
		\textsf{G}_{\textsf{ab}} = 8\pi G\expval{\textsf{T}_{\textsf{ab}}}{\Phi},
	\end{equation}
	where $\textsf{G}_{\textsf{ab}}$ is the Einstein tensor \cite{Birrell:1982ix}. These field equations reveal that $\expval{\textsf{T}_{\textsf{ab}}}$ acts as a source to gravity and variations in the matter sector backreact on the gravity sector. It is critical to the semiclassical approach to gravity to determine $\expval{\textsf{T}_{\textsf{ab}}}$ and ensure that its behavior does not dramatically affect the classical evolution. However, $\expval{\textsf{T}_{\textsf{ab}}}$ is an ill-defined operator valued distribution which is singular at a point. To control this divergence, we are lead to the Hadamard renormalization procedure. 

	In \cite{Wald:1977up}, Wald found that the renormalization of the stress--energy tensor is connected to the singular nature of the symmetrized two--point Wightman function, 
	\begin{equation}\label{eqn2:HEGF}
		\textsf{G}_\textsf{H}(x^\prime,x) = \tfrac{1}{2}\expval{\{\hat{\Psi}(x^\prime),\hat{\Psi}(x)\}}{\Phi}.
	\end{equation}
	The Wightman function satisfies the same equation of motion as the field, and captures the state-dependence of field correlations, assuming Gaussian states. 

	In \cite{Hadamard:1923}, Hadamard proved the ansatz that the Green's function for second order hyperbolic wave equations, such as $\square{\Psi}=0$, have a standard geometric divergence in the coincident limit. The existence of these Green's functions is a necessary condition to start considering a renormalization procedure. A sufficient condition, however, is to ensure that the singularity structure is preserved. It was shown that these conditions are preserved under Cauchy evolution in a hyperbolic spacetime and a large set of states exhibit the same divergence \cite{Fulling:1978ht,Fulling:1981cf}. Through point--splitting, these conditions manifest as a reliable renormalization procedure to obtain a finite value for $\expval{\textsf{T}_{\textsf{ab}}}$, 
	\begin{equation}
		\expval{\tilde{\textsf{T}}_{\textsf{ab}}} \sim \lim_{x^\prime\to x}\textsf{D}_{\textsf{a}\textsf{b}}(x^\prime,x)~\tilde{\textsf{G}}(x^\prime,x)
	\end{equation}
	where $\textsf{D}_{\textsf{a}\textsf{b}}(x^\prime,x)$ is a second order differential operator, $\tilde{\textsf{G}}(x^\prime,x)$ is the convergent piece of Eqn.~\eqref{eqn2:HEGF} at coincident limit, and the primed indices denote derivatives evaluated at the primed coordinates \cite{PhysRevD.34.2286,Decanini:2005gt,Decanini:2005eg}.

	Interestingly, it has been shown that the vacuum state of an inertial observer is a thermal state for an accelerated observer \cite{Unruh:1976db}. In fact, the temperature from such a thermal state can also be derived from the Green's function of a scalar field accelerating on a flat geometry. In \cite{Lynch:2016eje}, a generalized notion of the instantaneous temperature was developed from the Hadamard elementary Green's function and was linked to the local curvature, the acceleration, and the renormalized quantum vacuum polarization $\langle \tilde{\Psi}^2 \rangle$. However, $\langle \tilde{\Psi}^2 \rangle$ is not a physical observable in a massless minimally coupled theory, and thus derivative coupling could provide a more physical notion of temperature. Thus, in this paper, we propose a temperature that incorporates the effects of the derivative couplings of a generic quantum field theory. In this approach, $\langle \tilde{\Psi}^2 \rangle$ is replaced by $\expval{\textsf{T}_{\textsf{ab}}}$, which is a more physical observable and is well--defined for Hadamard states.

	The paper is organized as follows.\ In Section~\ref{sec:QFTinFS}, we present an outline of motivations.\ In Section~\ref{sec:GenTemp}, a method of building temperature via derivative couplings in flat spacetime is presented. In Section~\ref{sec:QFTCS}, we highlight the issues of quantizing a field in curved spacetime. Section~\ref{sec:HadaAnsatz} shows the development of the Hadamard ansatz. In Section~\ref{sec:thelaw}, we present the temperature registered by accelerated observers for generic Hadamard states in curved geometries. Finally, Section~\ref{sec:conclusion} concludes the paper and discusses further applications.

\section{Motivation: Quantum Field Theory in Flat Spacetime} % (fold)
	\label{sec:QFTinFS}
	In a four dimensional Minkowskian background, a massless scalar field, $\Psi(x)\equiv\Psi(\bm{x},t)$, freely propagates according to the wave equation,
	\begin{equation}
		\square_x\Psi(\bm{x},t)\equiv\qty(\pdv[2]{t}-\Delta_{\bm{x}})\Psi(\bm{x},t)=0
	\end{equation}
	where $\Delta_{\bm{x}}$ is the spatial Laplacian of the scalar field. If we wish to quantize the theory, the scalar field is promoted to an operator $\hat{\Psi}(\bm{x},t)$ satisfying the wave equation via a plane wave mode representation
	\begin{align}
		\hat{\Psi}(\bm{x},t)=&\int\frac{\dd^3\bm{k}}{(2\pi)^3}\frac{1}{\sqrt{2{\omega}}}\\
		&\times\qty{a_{\bm{k}}^\dagger e^{i({\omega}t-\bm{k}\cdot\bm{x})}+a^{{\color{white}\dagger}}_{\bm{k}}e^{-i({\omega}t-\bm{k}\cdot\bm{x})}},\nonumber
	\end{align}
	with the annihilation, $a^{{\color{white}\dagger}}_{\bm{k}}\ket{0}=\ket{0}$, and creation, $a^{\dagger}_{\bm{k}}\ket{0}=\ket{\bm{k}}$, operators satisfying the canonical commutation algebra of $[a^{{\color{white}\dagger}}_{\bm{k}},a^{{\dagger}}_{{\bm{k}^\prime}}]=(2\pi)^3\delta^3(\bm{k}-{\bm{k}^\prime}).$ Beyond the form of the field, the two--point Wightman function plays a central role as these functions describe the wavefunction of a Gaussian state. For inertial observers in Minkowski spacetime, the vacuum advanced Wightman function (assuming $t' > t$),
	\begin{align}
		\textsf{G}^+_{\textsf{M}}(\bm{x}^\prime,t^\prime;\bm{x},t)&\equiv\expval{\hat{\Psi}(\bm{x}^\prime,t^\prime)\hat{\Psi}(\bm{x},t)}{0}\\
		&=\int\frac{\dd^4 k}{(2\pi)^4}\frac{e^{-ik\cdot(x^\prime-x)}}{{\omega}^2-|\bm{k}|^2}
	\end{align}
	satisfies the wave equation, $\Box_x\textsf{G}_\textsf{M}^+=0$. Here, the $k_0=\omega$ integral is a contour integral around the pole ${\omega}=+|\bm{k}|$, and the Wightman function has the form
	\begin{equation}
		\textsf{G}^+_{\textsf{M}}(x^\prime,x) = -\frac{1}{4\pi^2}\qty[\frac{1}{(t^\prime-t-i\epsilon)^2-|\bm{x}^\prime-\bm{x}|^2}].
	\end{equation}

	Thus far, we have considered only inertial observers in a flat background. In order to step towards a quantum field theory in a curved background, we start by considering accelerated observers on a flat background. The simplest case is of observers with constant acceleration, $a$, moving along the $(t,x)$ plane in Minkowski spacetime,
	\begin{equation}
		y=z=0,\hspace{.1in}x=+\sqrt{t^2+a^{-2}},\hspace{.1in}t(\tau)=-\frac{1}{a}\sinh(a\tau)
	\end{equation}
	where $\tau$ is the proper time of the observer. Under these transformations, the wave equation and the mode decomposition for the scalar wave change accordingly. For our purposes, it is more instructive to understand the corresponding change in the {symmetrized} Wightman function which takes on the form
	\begin{equation}\label{eqn10}
		\textsf{G}^{+}_{\textsf{A}}(x^\prime,x)=-\qty[\frac{4\pi}{a}\sinh(\frac{a s(x^\prime,x)}{2})]^{-2}
	\end{equation}
	where $s[x(\tau^\prime),x(\tau)]=\tau^\prime-\tau$ is the proper distance. 

\section{Generalizing Temperatures}
	\label{sec:GenTemp}
	Along an accelerated trajectory, if one chooses to couple their motion to a particle detector, the detector would respond to the absorption of quanta at the rate of
	\begin{equation}
		\Gamma \propto \int\dd\Delta\tau\, e^{-i\Delta E(\tau^\prime-\tau)}\textsf{G}^{+}(x^\prime,x),
	\end{equation}
	where $E>E_g$ is the energy of the excited state and $E_g$ is the ground state energy \cite{Birrell:1982ix}.

	The Wightman function, calculated in Eqn.~\eqref{eqn10}, for a uniformly accelerating observer in a flat background coincides exactly with the Wightman function for an inertial observer immersed in a thermal bath with a detector response of
	\begin{equation}
		\Gamma_\textsf{A}\propto \frac{1}{e^{2\pi(E-E_g)/a}-1}
	\end{equation}
	at a temperature $k_\text{B} T = {a}/{2\pi}$. This is the Unruh effect \cite{Unruh:1976db}. Though the field is in its vacuum state, the acceleration provided by an external force excites the field causing the emission of quanta which provides resistance to the motion. Therefore, the external force has to do additional work to continue maintaining constant acceleration and the detector understands this tug--of--war as a thermal bath. However, the stress--energy transforms covariantly, $\expval{:\textsf{T}_\textsf{ab}:}{0}=\expval{:\textsf{T}_{\textsf{a}^\prime\textsf{b}^\prime}:}{0}=0$, implying both that the registered particles are fictitious and elucidating the issues present with maintaining a meaningful notion of vacuum in the presence of acceleration. Interestingly, another avatar of the same temperature appears in Hawking radiation \cite{Hawking:1974sw} suggesting that this phenomenon is more fundamental than first anticipated.

	We anticipate issues calculating the temperature for accelerated observers in \textit{curved} backgrounds. Instead of relying on the thermal response function, we can instead build temperatures by differentiating the Wightman function. For instance, consider the series expansion
	\begin{equation}\label{eqn13}
		\textsf{G}^+_\textsf{A}(x^\prime,x)=-\frac{1}{4\pi^2s^2}+\frac{a^2}{48\pi^2}-\frac{a^4s^2}{960\pi^2}+\mathcal{O}(s^4).
	\end{equation}
	By removing the divergent term, and recalling the definition for Unruh temperature, we can express the Wightman function as a power series in $k_\text{B}T$,
	\begin{equation}
		\tilde{\textsf{G}}^+_\textsf{A}(x^\prime,x)=\frac{(k_\text{B}T)^2}{12}-\frac{\pi^2s^2}{60}(k_\text{B}T)^4+\mathcal{O}(s^4).
	\end{equation}
	In this paper, we will consider the next order term in the expansion, given by Eqn.~\eqref{eqn13}, 
	\begin{equation}\label{eqn15}
		\frac{\pi^2}{30}[k_\text{B}T(\tau)]^{4} \equiv \lim_{x^\prime\to x}\qty[\dv{}{\tau}\dv{}{{{\tau}^\prime}}\tilde{\textsf{G}}^+_\textsf{A}(x^\prime,x)],
	\end{equation}
	matching the Stefan--Boltzmann law for scalar radiation. Thus, relying on the thermalicity of the Wightman function, we may generalize the process of extracting temperatures and, in principle, if one has an analytic expression for the Wightman function, then the procedure may be extended to a curved setting.
	% section QFTinFS (end)

\section{Quantum Field Theory in Curved Spacetime}
	\label{sec:QFTCS}
	Suppose $\Psi$ is a classical scalar field propagating on a standard four dimensional time--orientable globally hyperbolic spacetime $(\mathcal{M},g_{{a}{b}})$ with a Cauchy initial hypersurface. We consider the manifold to be without a boundary, i.e., $\partial\mathcal{M}=\varnothing$, and the metric signature to be $(-,+,+,+)$. The action for the system is
	\begin{equation}
		\mathcal{S}[\Psi,g]=-\frac{1}{2\kappa}\int\limits_{\mathcal{M}}\qty[g^\textsf{ab}\nabla_\textsf{a}\Psi\nabla_\textsf{b}\Psi+(m^2+\xi\textsf{R})\Psi^2]\epsilon
	\end{equation}
	where $m^2$ is the mass of the scalar field, $\kappa=8\pi Gc^{-4}$ is Einstein's constant, $\textsf{R}$ is the Ricci scalar, and $\epsilon$ is the volume form of the manifold which, for a set of charts, is given by $\sqrt{-\det(g)}\,\dd^4x.$ Here $\xi$ is a dimensionless factor which determines the interaction strength between the geometry and the field, with $\xi=0$ being the minimal coupling, and $\xi=(1/6)$ being conformal coupling. We hereby work in units where $G=c=\hbar=1.$

	From the action, one may obtain both the equations of motion for the field and the classical stress--energy tensor via the usual variation principle. Extremizing the variation of the action with respect to the field provides the Klein--Gordon equation of motion~\cite{Birrell:1982ix}
	\begin{equation}
		\textsf{L}[\Psi]=(\Box - m^2 - \xi\textsf{R})\Psi = 0.
	\end{equation}	

	Meanwhile, the classical stress--energy tensor is determined by varying the action with respect to the metric~\cite{Birrell:1982ix}
	\begin{align}
		\textsf{T}_\textsf{ab} =&~(1-2\xi)\Psi_{;\textsf{a}}\Psi_{;\textsf{b}}+\qty(2\xi-\frac{1}{2})g_\textsf{ab}\Psi^{;\textsf{c}}\Psi_{;\textsf{c}}\nonumber\\
		&-2\xi\Psi\Psi_{;\textsf{ab}}+2\xi g_{\textsf{ab}}\Psi\Box\Psi\nonumber	\\
		&+\xi\qty(\textsf{R}_{\textsf{ab}}-\frac{1}{2}g_{\textsf{ab}}\textsf{R}-\half g_{\textsf{ab}}m^2)\Psi^2
	\end{align}
	Since the action of general relativity is invariant under spatio--temporal diffeomorphisms, the classical stress--energy tensor is locally conserved;
	\begin{equation}
		\nabla^\textsf{a}(\textsf{T}_\textsf{ab})=0.
	\end{equation}
	If the field is conformally coupled to the geometry, i.e., $\xi=\frac{1}{6}$, and is massless, $m=0$, then the extra symmetry imposes restrictions on the trace of the classical stress--energy tensor, namely
	\begin{equation}
		\textsf{T}^\textsf{a}{}_\textsf{a}=0.
	\end{equation}
	A massless, minimally coupled scalar field has a classical stress--energy tensor dependent purely on its derivatives.

	While the procedure to obtain the stress--energy tensor is straightforward in the classical picture, quantization presents some immediate issues; a non-unique definition of a vacuum state, geometrical ambiguities in the action, breaking local energy conservation, anomalies in the trace, and a divergent stress--energy tensor \cite{Birrell:1982ix,Wald:1977up}.
	
	First, suppose we impose a canonical quantization procedure by promoting the field to an operator, $\Psi\to\hat{\Psi}$. After performing the usual mode decomposition, the vacuum is no longer a well defined ground state for observers as the Poincar\'e group is not in general a symmetry group. From this lack of symmetry, one cannot maintain the definition of positive frequency and negative frequency modes as, with sufficient acceleration, a detector would register a thermal bath instead of a vacuum state, as demonstrated in the prior section. Furthermore, the normal ordering procedure for operators also no longer holds. Meanwhile, unitary time evolution of states is not guaranteed unless one evokes global hyperbolicity. Lastly, though $\expval{\textsf{T}_{\textsf{ab}}}{\Phi}$ is a more objective probe, it is a non--local operator built out of the field and its derivatives which is singular in the coincident limit.

	To circumvent these issues, previous efforts \cite{Kay:1988mu,Radzikowski:1996pa,Castagnino:1984mk,Wald:1977up,Fulling:1978ht,Fulling:1981cf} have indicated that there exists a class of states, called Hadamard states, which preserve unitary evolution of operators, maintain mode signature, and provide a finite value for $\expval{\textsf{T}_{\textsf{ab}}}{\Phi}.$ However, these states are not a magic bullet, the cost of using them is locality in spacetime. Moreover, they preserve the singularity structure but do not remove the singularities themselves. Instead, one is forced to remove the divergences by employing point--splitting.

	Surprisingly, these states (locally) provide an explicit form for the symmetrized Wightman function \cite{Kay:1988mu} which one may use to build $\textsf{G}_\textsf{H}(x^\prime,x) = \frac{1}{2}\expval{\{\hat{\Psi}(x^\prime),\hat{\Psi}(x)\}}{\Phi_\textsf{H}}$, where $\ket{\Phi_\textsf{H}}$ indicates that the state is of Hadamard type. Now, instead of varying the effective action to obtain the renormalized stress--energy tensor, one obtains the point--splitted stress--energy tensor via
	\begin{equation}
		\expval{\textsf{T}_\textsf{ab}} = \lim_{x^\prime\to x}\textsf{D}_{\textsf{a}\textsf{b}}(x^\prime,x)~\textsf{G}_{\textsf{H}}(x^\prime,x)
	\end{equation}
	where $\textsf{D}_{\textsf{a}\textsf{b}}(x^\prime,x)$ is a second order differential operator defined at points $x$ and $x^\prime$ respectively via \cite{PhysRevD.34.2286,Decanini:2005eg},
	\begin{align}
		\textsf{D}_{\textsf{a}\textsf{b}} &=(1-2\xi)g_{\textsf{b}}{}^{\textsf{b}^\prime}\nabla_{\textsf{a}}\nabla_{\textsf{b}^\prime}+\qty(2\xi-\frac{1}{2})g_{\textsf{a}\textsf{b}}g^{\textsf{c}{\textsf{d}^{\prime}}}\nabla_{\textsf{c}}\nabla_{{\textsf{d}^{\prime}}}\nonumber\\
		&-2\xi\qty(g_{\textsf{a}}{}^{\textsf{a}^\prime}g_{\textsf{b}}{}^{\textsf{b}^\prime}\nabla_{\textsf{a}^\prime}\nabla_{\textsf{b}^\prime} - g_\textsf{ab}\nabla_\textsf{c}\nabla^{\textsf{c}})\nonumber\\
		&+\xi\qty(\textsf{R}_{\textsf{ab}}-\frac{1}{2}\textsf{R} g_{\textsf{a}\textsf{b}})-\frac{1}{2}m^2g_{\textsf{a}\textsf{b}}.
	\end{align}
	In what follows, we will embark on a discussion on the form of symmetrized Wightman function provided by the Hadamard ansatz. Along the way, we will make a few remarks on the convergent, and divergent, pieces of the ansatz and discuss the rest of the issues which arise from the quantization of $\textsf{T}_\textsf{ab}$, such as local energy conservation, and anomalies arising in the trace.

	By carefully treating the pathological behaviors in the stress--energy tensor, we build on the work made in \cite{Lynch:2016eje}, and reveal a connection between the temperature detected by an accelerated observer in a curved background and the renormalized stress--energy tensor.

	%As $\textsf{T}_\text{ab}$ is known to diverge, the authors of \cite{Wald:1977up,Kay:1988mu,Fulling:1978ht,Fulling:1981cf} illustrated the convergent part of the two--point function, $\tilde{\textsf{G}}_\textsf{H}(x^\prime,x)$, can be used to determine the finite piece of $\expval{\textsf{T}_\textsf{ab}}$ up to a local geometric contribution, that is
	%\begin{equation}
	%	\expval{\tilde{\textsf{T}}_\textsf{ab}} = \qty(\lim_{x^\prime\to x}\textsf{D}_{\textsf{a}\textsf{b}^\prime}(x^\prime,x)~\tilde{\textsf{G}}_\textsf{H}(x^\prime,x))+\tilde{\Theta}_{\textsf{ab}}(x).
	%\end{equation}
	%In the following sections, we extend the work of \cite{Lynch:2016eje} and reveal the connection between the temperature detected by an accelerated observer in a curved background and the renormalized stress--energy tensor through $\textsf{G}_\textsf{H}(x^\prime,x)$.

\section{The Hadamard Ansatz}
	\label{sec:HadaAnsatz}
	We will now explore the explicit form by assuming Hadamard's ansatz \cite{Hadamard:1923}. We assume that the scalar field has now been quantized and consider only Hadamard states, $\ket{\Phi_\textsf{H}}$. From the Klein-Gordon field equations, the symmetrized Wightman function satisfies
	\begin{equation}\label{eqn11:KGGF}
		(\Box_x-m^2-\xi \textsf{R})\textsf{G}_\textsf{H}(x^\prime,x) = 0
	\end{equation}
	where the Laplace--Beltrami operator is defined with respect to $x$. As the above is a hyperbolic PDE, we consider solutions of the Hadamard form 
	\begin{align}
		\textsf{G}_\textsf{H}(x^\prime,x) = \frac{1}{8\pi^2}\bigg[&\frac{\textsf{U}(x^\prime,x)}{\sigma(x^\prime,x)}+\textsf{W}(x^\prime,x)\\&
		+\textsf{V}(x^\prime,x)\ln|\sigma(x^\prime,x)|\bigg]\nonumber,
	\end{align}
	where each of the composite terms are symmetric, smooth, biscalar functions regular at coincidence, $x^\prime\to x$ \cite{Hadamard:1923}. We provide some expository notes on each of these functions in the proceeding sections.

	The above ansatz only holds for a local patch of spacetime where a field point $x^\prime\equiv x(\tau^\prime)$ is linked to the base point $x\equiv x(\tau)$ via a timelike curve parameterized by proper time $\tau$. Furthermore, the singular nature of the ansatz is apparent in both the $\log|\sigma|$ and $1/\sigma$ terms. Therefore, the purely finite part of the two--point Wightman function is given by
	\begin{equation}
		(8\pi^2)\,\tilde{\textsf{G}}_\textsf{H}(x^\prime,x) = \textsf{W}(x^\prime,x).
	\end{equation}
	Notice that deconvoluting the Klein-Gordon field equation actually gives rise to the Feynman Green's function, or the time--ordered product of $\hat{\Psi}(x)$ and $\hat{\Psi}(x^\prime)$. However, the real part of the Feynman Green's is the symmetrized two--point Wightman function which is more interesting as it satisfies the same differential equation for the operator $\hat{\Psi}(x)$ and is required to renormalize the stress--energy tensor. Lastly, in order to satisfy Eqn.~\eqref{eqn11:KGGF}, the operation
	\begin{align}
		8\pi^2 \Box_x \textsf{G}^{(1)} = &-\sigma^{-2}(2\textsf{U}_{;\textsf{a}}-\textsf{U}\Delta^{-1}\Delta_{;\textsf{a}})\sigma^{;\textsf{a}}\\
		&-\sigma^{-1}[2\textsf{V}+(2\textsf{V}_{;\textsf{a}}-\textsf{V}\Delta^{-1}\Delta_{;\textsf{a}})\sigma^{;\textsf{a}}+\Box_x\textsf{U}]\nonumber\\
		&+\Box_x\textsf{V}\ln|\sigma|+\Box_x\textsf{W}\nonumber
	\end{align}
	necessitates the following consistency conditions;
	\begin{align}
		(2\textsf{U}_{;\textsf{a}}-\textsf{U}\Delta^{-1}\Delta_{;\textsf{a}})\sigma^{;\textsf{a}}&=0\\
		(\Box_x-m^2-\xi\textsf{R})\textsf{V}&=0\\
		2\textsf{V}+(2\textsf{V}_{;\textsf{a}}-\textsf{V}\Delta^{-1}\Delta_{;\textsf{a}})\sigma^{;\textsf{a}}+&\nonumber\\
		(\Box_x-m^2-\xi\textsf{R})\textsf{U}+\sigma(\Box_x-m^2-\xi\textsf{R})\textsf{W}&=0.\label{eqn29}
	\end{align}
	These conditions are thoroughly explored in other work, primarily in \cite{DeWitt:1960fc,Decanini:2005gt,Decanini:2005eg}. In general, the exact forms of these functions are difficult to obtain.

	\subsection{The Synge's World Function \texorpdfstring{$\sigma(x^\prime,x)$}{Text}}
	We refer the interested reader to \cite{Poisson:2011nh} for an in--depth analysis of the world function. It suffices to say that the world function acts as a bridge \textit{uniquely} linking the quantum field, and the Green's function, at $x(\tau)$ to $x(\tau^\prime)$. 

	The Synge's world function, $\sigma(x^\prime,x),$ is a scalar function connects the base point, $x$, to the field point, $x^\prime$, within a neighborhood. Formally, the world function is half of the geodesic distance squared and satisfies
	\begin{equation}
		\frac{1}{2} g^{\textsf{a}^\prime\textsf{b}^\prime}\sigma_{;\textsf{a}^\prime} \sigma_{;\textsf{b}^\prime}= \frac{1}{2}g^{\textsf{ab}} \sigma_{;\textsf{a}}\sigma_{;\textsf{b}} = \sigma(x^\prime,x),
	\end{equation}
	where we denote tensors, and derivatives, at $x^\prime$ with primed indices. On timelike geodesics, this reduces to the identity $\sigma(x^\prime,x)=-s^2/2$. Additionally, the world function is symmetric and a scalar both at $x^\prime$ and $x$, and enjoys the following properties
	\begin{align}
		&\lim_{x^\prime\to x} \sigma(x^\prime,x) = 0,\\
		&\lim_{x^\prime\to x}\sigma_{;\textsf{a}^\prime\textsf{b}^\prime}=g_{\textsf{ab}},\\
		&\lim_{x^\prime\to x}\sigma_{;\textsf{b}^\prime}=0,\text{ and }\lim_{x\to x^\prime}\sigma_{;\textsf{a}}=0
	\end{align}
	
	This world function aids us to define the parallel propagator along the world line, $g_{\textsf{a}\textsf{b}^\prime}(x^\prime,x)$, which satisfies
	\begin{align}
		&(g_{\textsf{a}{\textsf{b}^{\prime}}}(x^\prime,x))_{;\textsf{c}}\,\sigma^{\textsf{c}}(x^\prime,x)=0,\\
		&g^{\textsf{a}}{}_{\textsf{b}^{\prime}}(x,x^\prime)\textsf{X}^{\textsf{b}^{\prime}}(x^\prime) = \textsf{X}^\textsf{a}(x), \\
		&g^{{\textsf{b}^{\prime}}}{}_{\textsf{a}}(x^\prime,x)\textsf{Y}^{\textsf{a}}(x) = \textsf{Y}^{{\textsf{b}^{\prime}}}(x^\prime),
	\end{align}
	where $\textsf{X}^{\textsf{b}^{\prime}}$ and $\textsf{Y}^\textsf{a}$ are tangent vectors at $x^\prime$ and $x$ respectively and $\lim_{x^\prime\to x}g_{\textsf{a}{\textsf{b}^{\prime}}}\to g_{\textsf{ab}}$. The parallel propagator transports vectors at $x^\prime$ to $x$ (and vice versa) along a unique geodesic provided by the world function.

	We conclude this section by discussing the quasilocal expansion \cite{Galley:2005tj}, Eqns.~(\ref{eqn24:Quasi1}--\ref{eqn25:Quasi2}), and various covariant derivatives of the world function, Eqns.~(\ref{eqn26:Dera1}--\ref{eqn28:Dera3});
	\begin{align}
		\sigma &=-\frac{s^2}{2}-\frac{s^4}{24}A^2-\frac{s^6}{720}A^4+\mathcal{O}(s^8),\label{eqn24:Quasi1}\\
		\sigma^{;\textsf{a}} &= -su^\textsf{a}-\frac{s^2}{2}a^\textsf{a}-\frac{s^3}{6}\dot{a}^\textsf{a}-\frac{s^4}{24}\ddot{a}^\textsf{a}+\mathcal{O}(s^5),\label{eqn25:Quasi2}\\
		\sigma^{;\textsf{a}\textsf{b}} &= g^{\textsf{ab}}-\frac{1}{3}\textsf{R}^{\textsf{a}~\textsf{b}}_{~\textsf{c}~\textsf{d}}\sigma^{;\textsf{c}}\sigma^{;\textsf{d}} +\mathcal{O}\qty({\sigma^{\frac{3}{2}}}),\label{eqn26:Dera1}\\
		\sigma^{;\textsf{a}{\textsf{b}^{\prime}}} &= -g_{\textsf{f}}^{~{\textsf{b}^{\prime}}}\qty(g^{\textsf{af}}-\frac{1}{6}\textsf{R}^{\textsf{a}~\textsf{f}}_{~\textsf{c}~\textsf{d}}\sigma^{;\textsf{c}}\sigma^{;\textsf{d}})+\mathcal{O}\qty(\sigma^{\frac{3}{2}}),~\&\label{eqn27:Dera2}\\
		\sigma^{;{\textsf{a}^{\prime}}{\textsf{b}^{\prime}}}&= g_{\textsf{e}}^{~{\textsf{a}^{\prime}}}g_{\textsf{f}}^{~{\textsf{b}^{\prime}}}\qty(g^{\textsf{ef}}-\frac{1}{3}\textsf{R}^{\textsf{e}~\textsf{f}}_{~\textsf{c}~\textsf{d}}\sigma^{;\textsf{c}}\sigma^{;\textsf{d}})+\mathcal{O}\qty({\sigma^{\frac{3}{2}}}),\label{eqn28:Dera3}
	\end{align}
	where the dot-notation denotes $\tau$ derivatives. Interestingly, the 4--velocity and 4--acceleration appear for the quantum field along the world line within the quasilocal expansion. The appearance of $A^2(x)=g_{\textsf{a}\textsf{b}}a^\textsf{a} a^\textsf{b}$ hints at a connection to the temperature of an accelerated observer on some curved background. 

	\subsection{Covariant Series Expansions} % (fold)
	\label{sub:covexpansions}
	The remaining functions within the Wightman function do not have exact forms. Instead, one has to employ a power series expansion for the functions 
	\begin{equation}
		\textsf{F}(x^\prime,x) = \sum_{n=0}^{+\infty} \textsf{F}_n(x^\prime,x)\sigma^{n}(x^\prime,x)
	\end{equation}
	and a covariant series expansion for each of the coefficients around the base point $x$,
	\begin{equation}
		\textsf{F}_n(x^\prime,x) = f_n(x) + \sum_{p=1}^{+\infty}\frac{(-1)^p}{p!}f_{n~a_1\cdots a_p}\sigma^{;a_1}\cdots\sigma^{;a_p}.
	\end{equation}
	Symmetric exchange of $x$ and $x^\prime$ and the wave equations form a set of constraints for the $p$--forms, $f_{a_1\cdots a_p}(x)$. These series expansions are the closest analogues to the Taylor series of flat space. Employing the covariant series expansions for the functions $\textsf{U}(x^\prime,x)$, $\textsf{V}(x^\prime,x)$, and $\textsf{W}(x^\prime,x)$ and imposing the consistency conditions at all orders in $s$ provides a recursive set of differential equations \cite{Decanini:2005eg}. In principle, if a boundary condition is provided, then one may \textit{exactly} reconstruct these functions. In practice, however, the difficulties of working in a curved setting makes this task impractical and, instead one could obtain an \textit{asymptotic} reconstruction up to the desired order. 

	From the covariant series expansions, the biscalars $\textsf{U}(x^\prime,x)$ and $\textsf{V}(x^\prime,x)$ have the form
	\begin{align}
		\textsf{U}(x^\prime,x)&=\textsf{U}_0(x^\prime,x)+\textsf{U}_1(x^\prime,x)\sigma(x^\prime,x)+\mathcal{O}(\sigma^2)\\
		&=u_0(x)-u_{0~\textsf{a}}\sigma^{;\textsf{a}}+\frac{1}{2}u_{0~\textsf{ab}}\sigma^{;\textsf{a}}\sigma^{;\textsf{b}}+\cdots\nonumber\\
		&+\sigma(x^\prime,x)(u_1(x)-u_{1~\textsf{a}}\sigma^{;\textsf{a}}+\cdots)+\mathcal{O}(\sigma^2)\nonumber
	\end{align}
	\begin{align}
		\textsf{V}(x^\prime,x)&=\textsf{V}_0(x^\prime,x)+\textsf{V}_1(x^\prime,x)\sigma(x^\prime,x)+\mathcal{O}(\sigma^2)\\
		&=v_0(x)-v_{0~\textsf{a}}\sigma^{;\textsf{a}}+\frac{1}{2}v_{0~\textsf{ab}}\sigma^{;\textsf{a}}\sigma^{;\textsf{b}}+\cdots\nonumber\\
		&+\sigma(x^\prime,x)(v_1(x)-v_{1~\textsf{a}}\sigma^{;\textsf{a}}+\cdots)+\mathcal{O}(\sigma^2)\nonumber
	\end{align}
	around the base point $x$ up to $s^2$. The boundary condition for $\textsf{U}(x^\prime,x)$ is given by
	\begin{equation}
		\textsf{U}_0(x^\prime,x)=\Delta^{\frac{1}{2}}(x^\prime,x)
	\end{equation}
	where the biscalar $\Delta(x^\prime,x)$ is the Van Vleck--Morette determinant 
	\begin{align}
		\Delta(x^\prime,x) &= -\frac{\det(-\sigma_{;\alpha\textsf{a}}(x^\prime,x))}{[-g(x)]^{1/2}[-g(x^\prime)]^{1/2}}
	\end{align}
	which roughly measures the direction one approaches the base point $x$ from the field point $x^\prime$ and satisfies $\lim_{x^\prime\to x}\Delta(x^\prime,x)=1$. The higher order terms, i.e., $\textsf{U}_n(x^\prime,x)$ for $n\ge1$, are not defined in four dimensions \cite{Decanini:2005gt}. Therefore, the only term which appears in the analysis is the boundary term, $\Delta^{\half}(x^\prime,x)$. 

	Meanwhile, the boundary condition for $\textsf{V}(x^\prime,x)$ is determined from the differential equation
	\begin{align}\label{eqn48}
		2\textsf{V}_0 &+ 2\textsf{V}_{0;\textsf{a}}\sigma^{;\textsf{a}} - 2\textsf{V}_0\Delta^{-1/2}\Delta^{1/2}{}_{;\textsf{a}}\sigma^{;\textsf{a}}\nonumber\\
		&+(\Box_{x}-m^2-\xi R)\Delta^{1/2}=0.
	\end{align}
	Much work \cite{Decanini:2005eg,Decanini:2005gt,DeWitt:1960fc} has been done to determine these coefficients to be used. We simply recite the relevant results from \cite{Decanini:2005eg,Decanini:2005gt} up to second order
	\begin{widetext}
		\begin{alignat}{2}
		\Delta^{\frac{1}{2}}(x^\prime,x)=&~1+\frac{\textsf{R}_{\textsf{a}\textsf{b}}}{12}\sigma^{;\textsf{a}} \sigma^{;\textsf{b}}-\frac{\textsf{R}_{\textsf{ab};\textsf{c}}}{24}\sigma^{;\textsf{a}} \sigma^{;\textsf{b}} \sigma^{;\textsf{c}}+\frac{1}{4!}\left[\frac{3}{10} \textsf{R}_{\textsf{ab};\textsf{cd}}+\frac{1}{15}\textsf{R}^{\textsf{f} }{}_{\textsf{a} \textsf{e} \textsf{b} } \textsf{R}^{\textsf{e}}{}_{\textsf{c} \textsf{f} \textsf{d} } +\frac{1}{12} \textsf{R}_{\textsf{ab}} \textsf{R}_{\textsf{cd} }\right]\sigma^{;\textsf{a}} \sigma^{;\textsf{b}} \sigma^{;\textsf{c}}\sigma^{;\textsf{d}} +\mathcal{O}(\sigma^{\frac{5}{2}}).\\
		\textsf{V}(x^\prime,x)=&~\sigma^{0}\times\bigg\{\frac{1}{2}\left[m^2+{\textsf{R}} \left(\xi -\frac{1}{6}\right) \right]-\frac{\textsf{R}_{;\textsf{a}}}{4}\left[\xi -\frac{1}{6}\right]\sigma^{;\textsf{a}}+\frac{\sigma^{;\textsf{a}} \sigma^{;\textsf{b}}}{2}\bigg[\frac{1}{12} \left(\xi -\frac{1}{6}\right) \textsf{R} \textsf{R}_{\textsf{a} \textsf{b} }+\frac{1}{6} \left(\xi -\frac{3}{20}\right) \textsf{R}_{;\textsf{a} \textsf{b} }\\
		&~~~~~~~~~~+\frac{ m^2 \textsf{R}_{\textsf{a} \textsf{b} }}{12}-\frac{1}{180}\textsf{R}^{\textsf{def} }{}_{\textsf{a}}\textsf{R}_{\textsf{def} \textsf{b}}-\frac{1}{180} \textsf{R}^{\textsf{de} } \textsf{R}_{\textsf{d} \textsf{a} \textsf{e} \textsf{b} }+\frac{1}{90}\textsf{R}_{\textsf{a}}{}^{\textsf{d}}\textsf{R}_{\textsf{d} \textsf{b}}-\frac{1}{120} \square_x \textsf{R}_{\textsf{a} \textsf{b} }\bigg]\bigg\}\nonumber\\
		+&~\sigma^{1}\times\bigg\{\frac{1}{8}\qty[m^2+\textsf{R}\qty(\xi-\frac{1}{6})]^2 - \frac{\square_x \textsf{R}}{24} \left(\xi -\frac{1}{5}\right)-\frac{1}{720}(\textsf{R}_{\textsf{de}}\textsf{R}^{\textsf{de}} - \textsf{R}^{\textsf{defg}}\textsf{R}_{\textsf{defg}}) \bigg\} + \mathcal{O}(\sigma^{\frac{3}{2}}).\nonumber
	\end{alignat}
	\end{widetext}
	
	In the process which we developed for a generalized notion of temperature, we removed the divergent piece of $\textsf{G}_\textsf{A}^+(x^\prime,x)$ before declaring the finite part of expression as the temperature and building temperatures at all orders. As $\textsf{U}/\sigma$ and $\textsf{V}\ln|\sigma|$ are clearly divergent in the coincident limit, the above expressions are necessary to control these divergences via point splitting. In the following section, we will motivate the remaining function in the Hadamard ansatz, $\textsf{W}(x^\prime,x)$, which is the prime candidate to contain the quantum state dependence.
	% subsection covariant_series_expansions (end)

	\subsection{The Quantum State \texorpdfstring{$\textsf{W}(x^\prime,x)$}{TEXT}}
	The interesting feature about this smooth symmetric biscalar lies in the freedom to specify a boundary condition. Though at first this seems to be a hindrance, the presence of boundary conditions for the other functions constrained them to be independent of the Hadamard state. The freedom provided by $\textsf{W}(x^\prime,x)$ allows one to encode the rest of the information for which a Wightman function is responsible, i.e., the quantum state. Therefore, the quantum state function, $\textsf{W}(x^\prime,x)$, is not only integral to ascertain a finite temperature from the two--point Wightman function, it connects the temperature with field operators, such as stress--energy tensor. 

\begin{figure}
		\includegraphics[width=\columnwidth]{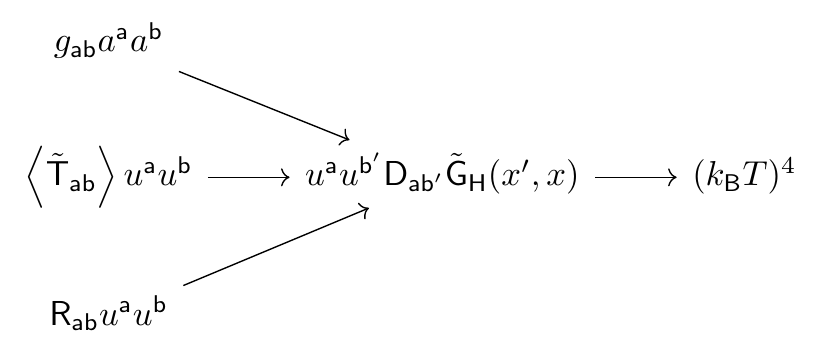}
		\caption{The above diagram illustrates the connection between the projected stress--energy tensor, among others, and $(k_\text{B}T)^4$ through the finite part of the Green's function.}
		\label{Fig1}
	\end{figure}

	In \cite{Lynch:2016eje}, the potential for the quantum state function to accurately reproduce the temperature for a massless, conformally coupled quantum field in de Sitter was demonstrated. There, they considered the zeroth order contribution from the convergent part of $\textsf{G}_\textsf{H}(x^\prime,x)$ and found,
	\begin{align}
		[k_\text{B}T_*(\tau)]^2 &= \lim_{x^\prime\to x}c_0~\tilde{\textsf{G}}^+_\textsf{H}(x^\prime,x)\\
		&=\frac{1}{(2\pi)^2}[A^2-\textsf{R}_{\textsf{ab}}u^\textsf{a}u^\textsf{b}+w_0(x)]\nonumber
	\end{align}
	after the covariant series for the coefficients of $\textsf{W}_n(x^\prime,x)$ was performed. Existing literature, for example \cite{PhysRevD.34.2286}, indicates that $w_0(x)$ is precisely the renormalized vacuum polarization, $w_0(x)=48\pi^2\langle{\tilde{\Psi}(x)^2}\rangle$. In the case of de Sitter, the above expression reduces to
	\begin{equation}
		(k_\text{B}T_\textsf{dS}(\tau))^2 = \frac{A^2+H^2}{(2\pi)^2}
	\end{equation}
	which is the expected temperature for observers in static de Sitter \cite{Obadia:2008rt,Narnhofer:1996zk}. In their non--kinetic case, the stress--energy tensor did not appear in the temperature at order $(k_\text{B}T_*)^2$ as the $\expval{\textsf{T}_\textsf{ab}}$ sees \textit{derivatives} of the two--point function. In fact, the massless, minimally coupled case, the stress--energy tensor is independent of the renormalized vacuum polarization. It seems the order which interesting physical connections appear between the temperature and the stress--energy tensor is at first derivative couplings of ${\textsf{G}}_\textsf{H}(x^\prime,x)$. This is not surprising since $(k_\text{B}T_*)^4$ has units of energy density and the differential operator $\textsf{D}_{\textsf{a}{\textsf{b}^{\prime}}}(x^\prime,x)$, used to build the renormalized $\expval{{\textsf{T}}_\textsf{ab}}$, is dependent on the derivative couplings of the finite part of $\textsf{G}_\textsf{H}(x^\prime,x)$, as shown in Fig.~\ref{Fig1}. 

	So there is a strong reason to suspect $\textsf{W}(x^\prime,x)$ is responsible for encoding not only the field's potential but also the field dynamics as well. These are precisely the arguments used in the Hadamard renormalization procedure. In fact, the first four Wald axioms \cite{Wald:1977up} renormalize
	\begin{align}\label{eqn53}
		\expval{\textsf{T}_\textsf{ab}(x)} &= \lim_{x^\prime\to x}~\textsf{D}_{\textsf{a}{\textsf{b}^{\prime}}}(x^\prime,x)~\textsf{G}_\textsf{H}(x^\prime,x)\longrightarrow\\
		\expval{\tilde{\textsf{T}}_\textsf{ab}(x)}&=\frac{\alpha_d}{2}\bigg[\lim_{x^\prime\to x}~\textsf{D}_{\textsf{a}{\textsf{b}^{\prime}}}(x^\prime,x)~\textsf{W}(x^\prime,x)\bigg]+{\Theta}_{\textsf{ab}}(x)\nonumber
	\end{align}
	where $\Theta_\textsf{ab}$ is state independent and $\alpha_d=1/(4\pi^2)$.

	Until now, the Hadamard ansatz for the Wightman function has been dominated by expansions which contain purely geometric contributions. As we have mentioned before, the geometric contributions couple to divergent functions indicating that $\textsf{W}(x^\prime,x)$ serves as a good candidate to renormalize $\expval{{\textsf{T}}_\textsf{ab}}$. However, when replacing the Wightman function with the quantum state function, it is not guaranteed that the conservation of $\expval{{\textsf{T}}_\textsf{ab}}$ implies the conservation of $\langle{\tilde{\textsf{T}}_\textsf{ab}}\rangle$ as the additional geometric contributions in $\textsf{G}_\textsf{H}(x^\prime,x)$ undoubtedly play a role in conserving $\expval{\textsf{T}_\textsf{ab}}$. Thus the renormalization procedure is valid up to a purely geometric, state--independent tensor $\Theta_\textsf{ab}$, which maintains energy conservation of the renormalized stress--energy tensor. The tensor, $\Theta_\textsf{ab}$ has to be state--independent as $\textsf{U}(x^\prime,x)$ and $\textsf{V}(x^\prime,x)$ are state--independent.

	Applying the covariant series expansion for $\textsf{V}_1(x^\prime,x)$ into the wave equation for $\textsf{W}(x^\prime,x)$ \cite{Decanini:2005eg}, one finds
	\begin{align}\label{eqn56}
		(\square_x&-m^2-\xi R)\textsf{W}(x^\prime,x) \\
		&= -6v_1(x)-2(v_{1}(x))_{;{\textsf{b}}}\sigma^{{\textsf{b}}}+\mathcal{O}(\sigma).\nonumber
	\end{align}
	Performing a power series expansion for $\textsf{W}(x^\prime,x)$ and a covariant series for its coefficients, \cite{Salehi:2000ev}, we find
	\begin{align}
		\textsf{W}(x^\prime,x) &= \textsf{W}_0(x^\prime,x)+\textsf{W}_1(x^\prime,x)\sigma(x^\prime,x)+\mathcal{O}(\sigma^2)\\
		&=w_0(x)-w_{0~\textsf{a}}\sigma^{;\textsf{a}}+\frac{1}{2}w_{0~\textsf{ab}}\sigma^{;\textsf{a}}\sigma^{;\textsf{b}}+\cdots\nonumber\\
		&+\sigma(x^\prime,x)(w_1(x)-w_{1~\textsf{a}}\sigma^{;\textsf{a}}+\cdots)+\mathcal{O}(\sigma^2)\nonumber
	\end{align}
	The properties of these tensors are throughly discussed in \cite{Decanini:2005eg,Salehi:2000ev}. The more convenient form for $\textsf{W}(x^\prime,x)$ is given by \cite{Decanini:2005eg}. Though we maintain the notation of \cite{Decanini:2005eg}, the explicit form of the tensor $w_\textsf{ab}(x)$ is provided in \cite{Salehi:2000ev}. From the wave equation for $\textsf{W}(x^\prime,x)$, we recite
	\begin{align}
		w_\textsf{ab}(x) &= \qty(w_{0~\textsf{ab}}-\frac{1}{4}g_{\textsf{ab}}w_{0}{}^{c}{}_{c})\nonumber\\
		&+\frac{g_{\textsf{ab}}}{4}\qty[(m^2+\xi \textsf{R})w(x)-6v_1(x)]\label{eqn58},\\
		w^\textsf{a}{}_{\textsf{c};\textsf{a}}(x) &= \frac{1}{2}\qty[\frac{1}{2}(\square w)_{;\textsf{c}}+\textsf{R}^\textsf{a}{}_\textsf{c}w_{;\textsf{a}}+\xi\textsf{R}_{;\textsf{c}}w-2{v_1}_{;\textsf{c}}]\label{eqn59},
	\end{align}
	and, from Eqn.~(\ref{eqn53}), $\langle\tilde{\textsf{T}}_\textsf{ab}\rangle$ has the form
	\begin{align}
		\expval{\tilde{\textsf{T}}_\textsf{ab}} &=\frac{\alpha_d}{2} \bigg[-\qty(w_{\textsf{ab}}-\half g_{\textsf{ab}}w^{\textsf{c}}{}_{\textsf{c}})\\
		&+\half(1-2\xi)w_{;\textsf{ab}}+\frac{g_{\textsf{ab}}}{2}\qty(2\xi-\half)\square w\nonumber\\
		&+\xi\qty(\textsf{R}_{\textsf{ab}}-\half g_{\textsf{ab}}\textsf{R}) w-\frac{g_{\textsf{ab}}}{2}m^2 w \bigg]+{\Theta}_{\textsf{ab}}.\nonumber
	\end{align}
	Using Eqn.~(\ref{eqn58}--\ref{eqn59}), $\nabla^\textsf{a}\expval{\tilde{\textsf{T}}_\textsf{ab}}=0$ provided
	\begin{equation}
		\qty[\Theta_\textsf{ab}-\alpha_d g_\textsf{ab}v_1(x)]^{;\textsf{a}}=0.
	\end{equation} 
	Integrating the above equation, one finds an equation for $\Theta_{\textsf{ab}}$ up to a locally conserved tensor, $\Omega_\textsf{ab}$,
	\begin{equation}
		\Theta_{\textsf{ab}}(x) = \alpha_d\, g_\textsf{ab} v_1(x) + {\Omega}_{\textsf{ab}}(x).
	\end{equation}
	The most general form that $\Omega_\textsf{ab}$ could have depends on the dimension of spacetime \cite{Decanini:2005eg}. In four dimensions, the tensor satisfying $\Omega_\textsf{ab}{}^{;\textsf{a}}=0$ is
	\begin{align}
		&\Omega_\textsf{ab} = A\,m^4 g_\textsf{ab} + B\, m^2[\textsf{R}_\textsf{ab}-(1/2)\textsf{R}g_{\textsf{ab}}]\\
		&+\frac{1}{\sqrt{-g}}\frac{\delta}{\delta g^{\textsf{ab}}}\int\limits_\mathcal{M}\qty[C_1\textsf{R}^2+C_2\textsf{R}_\textsf{ab}\textsf{R}^\textsf{ab}+C_3\textsf{R}_\textsf{abcd}\textsf{R}^\textsf{abcd}]\epsilon\nonumber
	\end{align}
	where $\epsilon$ is the volume form and $A$, $B$, $C_1$, $C_2$, and $C_3$ are arbitrary dimensionless constants with $C_i$ constrained by the Gauss--Bonnet Lagrangian \cite{Decanini:2005eg}. The locally conserved tensor has a trace \cite{Decanini:2005eg} of $g^{\textsf{ab}}\Omega_\textsf{ab} = -[6C_1+2C_2+2C_3]\square \textsf{R}$.
	
	Thus, we have a locally conserved stress--energy tensor which is connected to $w_{\textsf{ab}}$. The simplified form of this tensor is given by
	\begin{align}\label{eqn63}
		\expval{\tilde{\textsf{T}}_\textsf{ab}} =\frac{\alpha_d}{2} \bigg[&-w_{\textsf{ab}}+\half(1-2\xi)w_{;\textsf{ab}}-g_\textsf{ab}v_1\\
		&+\frac{g_{\textsf{ab}}}{2}\qty(2\xi-\half)\square w +\xi\textsf{R}_{\textsf{ab}} w\bigg]+{\Omega}_{\textsf{ab}},\nonumber
	\end{align}
	and with the trace being
	\begin{align}\label{trace}
		\expval{\tilde{\textsf{T}}^{\textsf{a}}{}_\textsf{a}} = \frac{\alpha_d}{2}\qty[3\qty(\xi-\xi_d)\square w-m^2w+2v_1] + \Omega^\textsf{a}{}_\textsf{a}
	\end{align}
	where $\xi_d=\frac{1}{4}\frac{d-2}{d-1}=\frac{1}{6}$ represents the conformal factor. Solving Eqns.~(\ref{eqn63}--\ref{trace}) for $w_\textsf{ab}$ and $\square w$ respectively, we find the following general expression
	\begin{align}
		&w_\textsf{ab}=\frac{2}{\alpha_d}\qty[\Omega_{\textsf{ab}}-\frac{g_{\textsf{ab}}}{2}\Omega^\textsf{c}{}_\textsf{c}-\qty(\expval{\tilde{\textsf{T}}_\textsf{ab}}-\frac{g_{\textsf{ab}}}{2}\expval{\tilde{\textsf{T}}^\textsf{c}{}_\textsf{c}})]\\
		&+\left[\frac{g_\textsf{ab}}{2}\qty({m^2w}-{\xi\square w})-4v_1+\xi \textsf{R}_\textsf{ab}w+\half(1-2\xi)w_{;\textsf{ab}}\right]\nonumber
	\end{align}
	which appears in the derivative couplings of the convergent part of the Green's function. As illustrated in Fig.~\ref{Fig1}, the state dependent tensor is projected along the observer's four--velocity.

	In four dimensions, when massless fields are conformally coupled, $\xi=\xi_d=\frac{1}{6},$ to gravity, the trace of the classical stress energy tensor vanishes. In the semiclassical case, labeling $v_1(x)\eval_{\xi_d}=v_c(x)$ we find instead 
	\begin{equation}
		\expval{\tilde{\textsf{T}}^{\textsf{a}}{}_\textsf{a}} = \alpha_d v_c(x) + g^{\textsf{ab}}\Omega_\textsf{ab}.
	\end{equation}
	In general, $g^\textsf{ab}\,\Omega_\textsf{ab}$ is proportional to $\square R$ \cite{Birrell:1982ix,Decanini:2005eg} and can be removed. On the other hand, $\alpha_dv_c(x)$ is a purely geometrical quantity which does not identically vanish. The presence of the trace anomaly is due to the key difference between the equation of motion for the quantum field,
	\begin{equation}
		(\square_x-m^2-\xi R)\Psi(x) = 0
	\end{equation}
	and the wave equation for the quantum state function Eqn.~(\ref{eqn56}), primarily in the presence of $-6v_1(x)$. 

	In order to ascertain the behavior of the trace anomaly and how it will influence the calculation of temperature, let us consider an interesting analogy. Consider the classical stress--energy tensor of a perfect fluid \cite{Carroll:2004st},
	\begin{equation}
		\textsf{T}^\textsf{ab} = (\rho+p)u^\textsf{a}u^\textsf{b}+pg^\textsf{ab}
	\end{equation}
	with trace $\textsf{T}^\textsf{a}{}_\textsf{a}=-\rho+3p$. For fluid--like vacuum energies, $p_{vac}=-\rho_{vac}$, one finds a trace of $\textsf{T}^\textsf{a}{}_\textsf{a} = 4p$. Returning to the renormalized quantum stress--energy tensor, if we interpret $2 v_c(x)=p_{a}(x)$ as an \textit{anomaly pressure}, the wave equation for $\textsf{W}(x^\prime,x)$ takes on an interesting form
	\begin{align}
		(\square_x&-m^2-\xi R)\textsf{W}(x^\prime,x)\approx -3p_a(x)
	\end{align}
	at zeroth order in $\sigma.$ Performing the covariant series expansion for $\textsf{W}(x^\prime,x)$ turns Eqn~(\ref{eqn58}) into 
	\begin{equation}
		w^\textsf{a}{}_\textsf{a}(x) = (m^2+\xi\textsf{R})w(x)-3p_a(x).
	\end{equation}
	This looks similar to the trace of a perfect fluid for an energy density of $\rho_w(x)=(m^2+\xi\textsf{R})w(x).$ Provided the analogy holds, $w_\textsf{ab}$ acts as the classical stress--energy tensor for the quantum field in the thermodynamic limit, and the renormalized quantum stress--energy tensor acts as a container for the anomaly pressure. Though in the massless, conformally coupled case, the trace is supposed to vanish, the quantum stress--energy tensor maintains some of the total anomaly pressure, $P_\textsf{T}(x)=3p_a(x)$,
	\begin{equation}
		\expval{\tilde{\textsf{T}}^\textsf{a}{}_\textsf{a}} = \frac{\alpha_d}{2}\frac{P_\textsf{T}(x)}{3}.
	\end{equation}
	Since classically $p\propto T^4$, we need to account for the remaining total pressure which is missing to obtain an accurate temperature reading. In the following section, we propose a solution to the missing pressure problem which works for all couplings of the field to the geometry, we correct the trace anomaly, and present the temperature registered by accelerated observers in curved geometries.

\section{Temperatures of Accelerated Observers in Curved Spacetime}
	\label{sec:thelaw}
	In the previous section, we developed a picture to understand the behavior of the trace anomaly for a massless, conformally coupled scalar field in a curved background. The analogy we constructed hints that the semiclassical stress--energy tensor contains some gravitational energy which is present in the trace at all levels of coupling between matter and gravity. In some ways, this is reminiscent of ultraviolet divergences.

	%In the thermodynamic limit, a massless scalar field conformally coupled to the geometry could also be treated as a perfect fluid with an effective mass being the coupling strength. The trace of the classical stress--energy tensor is then $3p-\rho=0$. This is the same equation satisfied for a black body in thermal equilibrium with its surroundings at temperature, $T$, experiencing a compressive pressure.

	%In standard Minkowski quantum field theory, normal ordering the Hamiltonian removes the ultraviolet divergences one finds due to the vacuum. Although vacuum effects should also vanish in the semiclassical case, since we do not have a reliable normal ordering procedure for the stress--energy tensor which contains the Hamiltonian density, we cannot expect the vacuum effects to vanish. If we naively combine the assumptions for a perfect fluid and vacuum energy density, then we would find $p_{vac}=-\rho_{vac}/3$. Now, holding $p_a(x)=2v_c(x)$, the previous analogy directly leads us to 
	%\begin{equation}\label{eqn71}
	%	\expval{\tilde{\textsf{T}}^\textsf{a}{}_\textsf{a}} \sim \frac{\alpha_d}{2}\frac{P_\textsf{T}(x)}{3}=-\frac{\alpha_d}{2}\frac{\rho_a(x)}{3}\sim-\frac{4\sigma}{3}T^4.
	%\end{equation}
	%If the analogy holds, the total anomaly pressure, $P_\textsf{T}$, is related to the internal radiant energy per unit \textit{volume} of the vacuum and the vacuum acts like a black body at thermal equilibrium at a temperature $T$ experiencing an \textit{expansive} pressure.

	Implicitly, we have been studying our quantum field theory in a local patch of spacetime using Hadamard states. This is akin to placing the quantum field theory in a box and only considering energy density in order to remove the infrared divergences. Since the diagonal of the stress--energy tensor contains the energy density, we have escaped the issues of an infrared divergence. However, the explicit form of the trace anomaly in the massless, conformally coupled case is
	\begin{equation}
		v_c(x) = \frac{1}{720}[\square_x\textsf{R}-\textsf{R}_\textsf{ab}\textsf{R}^\textsf{ab}+\textsf{R}_\textsf{abcd}\textsf{R}^\textsf{abcd}]
	\end{equation}
	which diverges in the high curvature limit. That is, the trace anomaly is a type of gravitational zero--point energy of spacetime.

	It is interesting to note that one indeed recovers the classical behavior from the quantum stress--energy tensor in the flat space limit. If one is in a nearly flat spacetime, then $\alpha_dv_c(x)\approx\lambda_z$, i.e., a constant zero--point energy contribution. In principle, the trace anomaly might appear as a contribution to the cosmological constant in any nearly flat solution of Einstein equations.

	It seems that our renormalization scheme is broken in the high curvature limit. This is not surprising because the semiclassical approximation is not a panacea and one requires a full theory of quantum gravity to circumvent these issues. With the presence of these divergences, the power of making thermodynamic statements is also lost. The analogy constructed thus far holds only in near equilibrium circumstances, i.e., when it is possible to define Hadamard states and a locally well behaved quantum field theory. The analogy attempts to satisfy Wald's fifth axiom \cite{Wald:1977up}, which demands a smooth transition from semiclassical behavior to the classical dynamics, namely the existence of a thermodynamic limit. 

	%As Eqn.~(\ref{eqn71}) indicates, the constant could be interpreted as the temperature of the vacuum at thermal equilibrium. 

	To adhere strictly to the fifth axiom, the massless, conformally coupled limit of the quantum stress--energy tensor needs to have a vanishing trace. The analogy constructed thus far provides a convenient ansatz, let $\textsf{A}_\textsf{ab}$ be an anomaly tensor of the perfect fluid form
	\begin{equation}
		\textsf{A}_\textsf{ab} = -\lambda(m,\xi;x)u_\textsf{a}u_\textsf{b}-\eta(m,\xi;x)(u_\textsf{a}u_\textsf{b}+g_\textsf{ab}),
	\end{equation}
	where $\lambda(m,\xi;x)$ and $\eta(m,\xi;x)$ are smooth functions which depend on the mass, coupling, and the geometry. The four--velocities here are assumed to be time--like for a family of observers.

	In order to maintain both energy conservation of $\langle\tilde{\textsf{T}}_\textsf{ab}\rangle$ and the relationship between $w_\textsf{ab}$ and $\langle\textsf{T}_\textsf{ab}\rangle$, we construct an auxiliary tensor, $\textsf{S}_\textsf{ab}$, which agrees with the classical stress--energy tensor in the massless, conformally coupled limit via
	\begin{equation}
		\textsf{S}_\textsf{ab} = \expval{\tilde{\textsf{T}}_\textsf{ab}} - \frac{\alpha_d}{2}\textsf{A}_\textsf{ab}.
	\end{equation}
	The freedom to construct $\textsf{S}_\textsf{ab}$ can be interpreted as a boundary condition being introduced on the derivatives of $\textsf{W}(x^\prime,x)$ or as a modification of the Hadamard renormalization procedure to include the fifth axiom.

	Now, in the massless, conformally coupled case, $\textsf{S}_\textsf{ab}$ can be made traceless to agree with the classical picture, provided the following condition holds
	\begin{align}
		\alpha_d v_c(x)-\frac{\alpha_d}{2}(\lambda(\xi,m;x)-3\eta(\xi,m;x))=0.
	\end{align}
	This provides one equation for two unknown functions. The second constraint is provided if we make contact with a relevant spacetime, such as de Sitter.

	Every horizon is accompanied with a temperature. In maximally symmetric de Sitter, the temperature \cite{Narnhofer:1996zk,PhysRevD.34.2286} is believed to be $k_\text{B}T_\textsf{dS} = H/2\pi$, where $H$ is the cosmological horizon. %The trace anomaly takes on an interesting form, $v_c(x) = -H^4/60$. This is perfectly in line with the equation of state for accelerating spacetimes $w = \frac{P_\textsf{T}}{\rho} = -1.$ Moreover, since $v_c(x) < 0$, $P_\textsf{T}<0$, and as a consequence $\rho_a(x)>0$. We are saved from the embarrassment of dealing with a negative energy densities.
	Demanding that our procedure provide us with the correct de Sitter temperature, we find 
	\begin{align}
		\lambda(m,\xi;x) &= \frac{1}{3}(v_c(x)-v_1(x)),\ \& \\
		\eta(m,\xi;x) &= -\frac{2}{3}v_c(x),
	\end{align}
	where $\lambda(m,\xi;x)$ vanishes in the massless, conformally coupled limit. The presence of $v_1(x)$ in $\lambda(m,\xi;x)$ arrived from the requirement that pure de Sitter temperature is the same temperature regardless of the coupling. Note that we project $\textsf{S}_\textsf{ab}$ along time--like four--velocities when we calculate the temperature.

	We are now in a position to present the temperature for accelerated observers in curved geometries. We first perform the covariant series expansion, found in Section~\ref{sub:covexpansions}, on the entire two--point function, $\textsf{G}_\textsf{H}(x^\prime,x)$. Since these expansions are asymptotic, it is crucial to perform the derivatives before applying the quasilocal approximation for $\sigma(x^\prime,x)$. At the relevant orders, some terms of $\textsf{U}(x^\prime,x)/\sigma(x^\prime,x)$ and $\textsf{V}(x^\prime,x)\ln|\sigma(x^\prime,x)|$ become finite after derivatives and appear in $\tilde{\textsf{G}}_\textsf{H}(x^\prime,x)$. After point--splitting the remaining divergent terms, one can take the coincident limit. This rather technical procedure is compactly represented in
	\begin{align}\label{thelaw}
		(k_\text{B}T_*(\tau))^4&\equiv\frac{30}{\pi^2}\lim_{x^\prime\to x}\qty[u^\textsf{a}u^{\textsf{b}^\prime} \nabla_{\textsf{a}}\nabla_{\textsf{b}^\prime}{\mathcal{G}}]\\
		&=\frac{1}{16\pi^4}(\textsf{D}(m,\xi;x)+\textsf{A}(m,\xi;x)+\textsf{G}(m,\xi;x))\nonumber
	\end{align}
	where ${\mathcal{G}}(x^\prime,x)$ reflects these changes, and the functions are shown below.
	\begin{figure*}[t]
	\includegraphics[width=\textwidth]{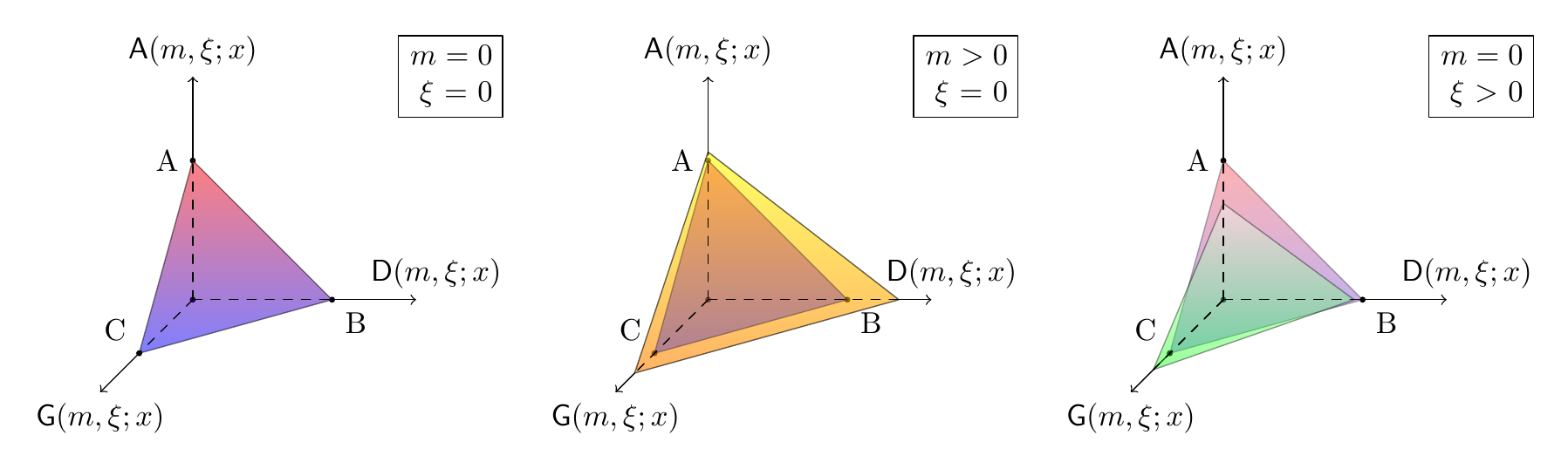}
	\caption{The three surfaces above characterize the temperature profile in parameter space for a variety of mass and coupling strengths. The points A, B, and C refer to Rindler, Thermal Minkowski, and to de Sitter space respectively.}
	\label{Fig2}
	\end{figure*}
	\begin{widetext}
		\begin{align}
			\textsf{D}(m,\xi;x)&=240\pi^2\qty[\expval{\tilde{\textsf{T}}^\textsf{a}{}_\textsf{a}}+2\expval{\tilde{\textsf{T}}_\textsf{ab}}u^\textsf{a}u^\textsf{b}]-30\xi(\square w-m^2w-2w_{;\textsf{ab}}u^\textsf{a}u^\textsf{b}+2\textsf{R}_\textsf{ab}u^\textsf{a}u^\textsf{b}w)\\
			\textsf{A}(m,\xi;x)&=A^4+5A^2\qty[\textsf{R}\qty(\frac{1}{6}-\xi)-\frac{1}{3}\textsf{R}_\textsf{ab}u^\textsf{a}u^\textsf{b}-m^2]+\frac{5}{2}a^\textsf{a}\qty[\textsf{R}_{;\textsf{a}}(1-6\xi)-4u^\textsf{b}u^{\textsf{c}}\textsf{R}_{\textsf{ab};\textsf{c}}-{2}a^\textsf{b}\textsf{R}_\textsf{ab}]-\frac{10}{3}u^\textsf{a}\dot{a}^\textsf{b}\textsf{R}_\textsf{ab}\\
			\textsf{G}(m,\xi;x)&=60v_1(x)-60\textsf{A}_\textsf{ab}u^\textsf{a}u^\textsf{b}-240\pi^2(\Omega^\textsf{a}{}_\textsf{a}+2\Omega_\textsf{ab}u^\textsf{a}u^\textsf{b})+{u}^\textsf{a}{u}^\textsf{b}{u}^\textsf{c}{u}^\textsf{d}\qty(\frac{2}{3}\textsf{R}^\textsf{e}{}_\textsf{afb}\textsf{R}^\textsf{f}{}_\textsf{ced}-2\textsf{R}_{\textsf{ab};\textsf{cd}}+\frac{5}{6}\textsf{R}_\textsf{ab}\textsf{R}_\textsf{cd})\nonumber\\
			&+u^\textsf{a}u^\textsf{b}\qty[\frac{5}{2}\textsf{R}\textsf{R}_\textsf{ab}(1-6\xi)-15 m^2\textsf{R}_\textsf{ab}-2\textsf{R}^\textsf{c}{}_\textsf{a} \textsf{R}_\textsf{cb}+\textsf{R}^\textsf{cd}\textsf{R}_\textsf{cadb}+\frac{3}{2}{\square \textsf{R}}_\textsf{ab}+\textsf{R}^\textsf{cde}{}_\textsf{a}\textsf{R}_\textsf{cdeb}-\frac{1}{2} \textsf{R}_{;\textsf{ab}}]
		\end{align}
	\end{widetext}

	The functions $\textsf{A}(m,\xi;x),$ $\textsf{D}(m,\xi;x)$, and $\textsf{G}(m,\xi;x)$ describe the acceleration, state dependence, and geometry contributions to the temperature respectively. In Fig.~\ref{Fig2}, the behavior of these functions for various couplings and masses is given. The above temperature definition agrees precisely with 4D Rindler, de Sitter, and thermal Minkowski discussed below.

	As an independent check of our results, we consider the case of a massless scalar field, minimally coupled to gravity in de Sitter. Here the de Sitter two--point function \cite{PhysRevD.34.2286} has the following Hadamard development 
	\begin{equation}
		\textsf{G}_{\textsf{dS}}(x^\prime,x) = \frac{\textsf{R}}{24\pi}\qty[\frac{1}{1-z}-\ln|1-z|-\ln(2t^\prime t)]
	\end{equation}
	where with $z(x^\prime,x)=\cos\sqrt{(\textsf{R}\sigma/6)}$, $w(x)=-\textsf{R}/6\ln(t)$, $\Omega_\textsf{ab}=\bm{0}$, and $\square_xw(x)=-\textsf{R}/24$. Here $\Delta(x^\prime,x)$, $\textsf{V}(x^\prime,x)$, and $\textsf{W}(x^\prime,x)$ are known exactly, the prior expansions are not required and the quantum stress--energy tensor has not been corrected. Straightforwardly performing $\tau,$ and $\tau^\prime$ derivatives of $\textsf{G}_{\textsf{DS}}(x^\prime,x)$ produces a quantity of
	\begin{equation}\label{eqn82}
		\frac{1}{16\pi^4}\qty(A^4+10H^2A^2+11H^4).
	\end{equation}
	Comparing this value to our definition, Eqn.~\eqref{thelaw}, in the de Sitter limit for accelerated observers, we find
	\begin{equation}\label{eqn83}
		(k_\text{B}T_{\textsf{dS}})^4\eval_{{m=0},~{\xi=0}}=\frac{1}{16\pi^4}\qty(A^4+10H^2A^2+H^4).
	\end{equation}
	The difference between Eqns.~(\ref{eqn82}--\ref{eqn83}) is $-10H^4/(16\pi^4)$ which is precisely the contribution from the anomaly tensor, specifically from $\lambda(m,\xi;x)$. For a massless scalar field, conformally coupled to gravity, we find
	\begin{equation}
		(k_\text{B}T_{\textsf{dS}})^4 \eval_{{m=0},~{\xi=1/6}}=\frac{1}{16\pi^4}(A^4+H^4).
	\end{equation}
	For Rindler, one recovers $T_\textsf{R}^4 = A^4/{16\pi^4}$ if one sets the mass, coupling, and the curvature to vanish. In Thermal Minkowski, for a massless, minimally coupled field, an inertial observer registers a temperature of
	\begin{align}
		(k_\text{B}T_{\textsf{TM}})^4 &= \frac{30}{\pi^2}\expval{\tilde{\textsf{T}}_{\textsf{ab}}}u^\textsf{a}u^\textsf{b}\nonumber\\
		&=-\frac{15}{4\pi^4}\qty(w_{0~\textsf{ab}}u^\textsf{a}u^\textsf{b}+\frac{1}{4}w_{0}{}^{\textsf{c}}{}_{\textsf{c}})\label{eqn85}
	\end{align}
	where we have used Eqns.~(\ref{eqn58},~\ref{eqn63}). In de Sitter, this contribution would vanish. However, in Thermal Minkowski, the first line reproduces the temperature of a thermal state and constrains the state dependence.

	%Lastly, it is interesting to note the presence of the trace anomaly and other purely geometric contributions appearing in $\textsf{G}(x,\xi)$ indicating a surviving gravitational zero--point energy. As observers only measure energy differences, these terms should not play a role. However, as gravity couples to all energies, neglecting these geometric contributions results in erroneous temperatures.

	\section{Conclusions} % (fold)
	\label{sec:conclusion}
	Based on our analysis, we present a generalized Stefan--Boltzmann law for spacetime,
	\begin{equation}
		j_* = \frac{\pi^2}{30} (k_\text{B}T_*)^4
	\end{equation}
	constructed from the result of Eqn.~\eqref{thelaw}. The temperature tetrahedron, Fig.~\ref{Fig2}, represents the calibration of derivative couplings of the Wightman function  at different limits. It is important to note that this definition can be used away from perfect thermal equilibrium in the presence of the non-trivial geometry $\textsf{G}(m,\xi;x)$, acceleration $\textsf{A}(m,\xi;x)$, and the state dependence $\textsf{D}(m,\xi;x)$. 

	We precisely recover the Stefan--Boltzmann factor of $\pi^2/30$ for scalar radiation previously mentioned in Eqn.~\eqref{eqn15}. In the Thermal Minkowski case, the four--velocities in Eqn.~\eqref{eqn85} can be chosen in such a way they represent the radiating energy density of a black body. In the Rindler case, our temperature definition reduces to the Unruh temperature, as desired. In the de Sitter case, after we have introduced the anomaly tensor, the temperature reads $k_\text{B}T_{\textsf{dS}} =H/2\pi$ at minimal and conformal couplings in the absence of mass and acceleration.

	In a first, the generality of these results opens new possibilities to explore dynamical solutions of the Einstein field equations and the existence of a zero--point gravitational energy. Our definition of temperature is built from the stress--energy tensor, and connects the semiclassical and classical picture in line with Wald's fifth axiom \cite{Wald:1977up}. One may use this temperature to study the thermodynamical nature of gravitational systems given suitable definitions of entropy, and other variables of state. Furthermore, this temperature definition could be used to probe thermal behavior and stability of spacetimes at the next highest order in the derivative expansion of the Wightman function. This method to construct temperatures could also be extended to higher orders.

	There exists a consensus in the literature that spacetimes have a thermodynamical interpretation \cite{Jacobson:1995ab}. For future research, it would interesting to explore the consequences of implementing our definition in the laws of spacetime thermodynamics.
	% section conclusions (end)

	\begin{acknowledgments} % (fold)
%	\label{sec:acknowledgments}
	We would like to thank Natacha Altamirano, Andrei Frolov, Soham Mukerjee, Eric Poisson, Matthew Robbins, and Rafael Sorkin for their valuable comments and discussions throughout the realization of this paper. Research at the Perimeter Institute is supported by the Government of Canada through the Department of Innovation, Science and Economic Development Canada. AD is supported by the National Science Foundation Graduate Research Fellowship Program under Grant No.\ 00039202. The work of JG and NA is partly funded by the Discovery Grants program of the Natural Sciences and Engineering Research Council of Canada. JG is further supported by the Billy Jones Scholarship granted by the Physics Department at Simon Fraser University.
	\end{acknowledgments}

	% section acknowledgments (end)

	\bibliography{DerivativeCouplings.bib}

	%{
	%	\bibliographystyle{unsrt}
	%	\bibliography{DerivativeCouplings.bib}
	%}

\end{document}